# Quantum entanglement and spin control in silicon nanocrystal

Vesna Berec [1, 2]

(1) Faculty of Physics, Department of Condensed Matter Physics,
University in Belgrade,
(2) Institute of Nuclear Sciences, Vinča,
P. O. Box 522, 11001 Belgrade, Serbia

## Abstract

Selective coherence control and electrically mediated exchange coupling of single electron spin between triplet and singlet states using numerically-derived optimal control of proton pulses is demonstrated. We obtained spatial confinement below size of the Bohr radius for proton spin chain FWHM. Precise manipulation of individual spins and polarization of electron spin states are analyzed via proton induced emission and controlled population of energy shells in pure $^{29}$Si nanocrystal. Entangled quantum states of channeled proton trajectories are mapped in transverse and angular phase space of $^{29}$Si $\langle 100 \rangle$ axial channel alignment in order to avoid transversal excitations. Proton density and proton energy as impact parameter functions are characterized in single particle density matrix via discretization of diagonal and nearest off-diagonal elements. We combined high field and low densities (1 MeV/92 nm) to create inseparable quantum state by superimposing the hyperpolarizationed proton spin chain with electron spin of $^{29}$Si. Quantum discretization of density of states (DOS) was performed by the Monte Carlo simulation method using numerical solutions of proton equations of motion. Distribution of gaussian coherent states is obtained by continuous modulation of individual spin phase and amplitude. Obtained results allow precise engineering and faithful mapping of spin states. This would provide the effective quantum key distribution (QKD) and transmission of quantum information over remote distances between quantum memory centers for scalable quantum communication network. Furthermore, obtained results give insights in application of channeled protons subatomic microscopy as a complete versatile scanning-probe system capable of both quantum engineering of charged particle states and characterization of quantum states below diffraction limit linear and in-depth resolution.

# Introduction

Major progress of experimental techniques as well as theoretical models during the last few decades, has made possible the comprehensive analysis of the ion beams collision dynamics [1, 2]. Obtained results have facilitated development of versatile analytical instruments which can provide material characterization, modification and analyses [3, 4] over a wide range of scientific disciplines. In addition, focused ion beam techniques beyond sub-nanometer scale [5-9] have gained an important role as silicon based nano-domain engineering [10, 11, 12] has become one of the most important tool in materials research, low dimensional system electronics, semiconductor manufacturing and nanotechnology overall. Recent experimental investigations of quantum information processing via single electron devices in gate defined quantum dots [13, 14] confirm silicon based spin quantum-information processor as a promising candidate for future quantum computer architectures [15]. In that context series of investigations of electrically [16, 17, 18] and optically [19] induced ion kinetics in solid state quantum systems reveal that focusing of coherent ions through oriented crystal, may enhance precise confinement and manipulation of individual spins in quantum information processing [20, 21]. The most prominent recent results relating the spin dynamics control to ion channeling techniques in thin crystals presented in series of theoretical studies [22-26] when the ion differential cross section is singular [27] give opportunity for precise manipulation of intrinsic properties of charged particles. In this paper we present theoretical study of localization and coherent control by superfocused channeled protons, CP beam induced polarization of individual electron spins in pure $^{29}$Si nanocrystal. We analyze precise control of entangled proton trajectories and discrete quantum states of phase space in connection to selective spin manipulation. The harmonic motion of highly correlated channeled protons is tuned by external RF field, by varying the CP energy and tilt angle relative to main $\langle 100 \rangle$ crystal axis. The calculations include the quasiharmonic approximation as well as the effect of multiple scattering by valence electrons and assume the anharmonicity of the interaction potential [28, 29]. Quantum entanglement of focused ion trajectories in final states corresponds to central part of the $\langle 100 \rangle$ Si axial channel. It is analyzed in phase space by convoluted transfer matrix method [30].

According to Liouville's theorem [31], the ensemble of channeled particles (for large impact parameters) experiences series of correlated, small angle collisions in initial stage of elastic interaction with atoms of the crystal lattice. Therefore, a proton flux distribution can be calculated via probability function of quantum trajectory reversibility, i.e., the probability for appearing of backscattered particles along initial propagation direction. The resultant flux distribution further considers unnormalized probability map of trajectories of channeled particles in phase space. We have analyzed the nonequilibrium state of channeled protons density profiles in configuration and scattering angle plane in connection to anharmonic expansion terms of proton - crystal effective potential. Calculation assumes the initial state of static equilibrium, considering 92 nm crystal's length and channeling conditions which correspond to infinitesimal crystal tilt angles, from zero up to 20% of critical angle for channeling. Degree of correlation between separate trajectories of channeled protons was calculated by two separate mapping procedures between configuration and angular phase plane. Thus, the nonharmonic-higher order terms of continuum interaction potential were analyzed via distribution function of channeled protons in transverse position plane and scattering angle plane.

The subsequent parts of paper are organized as follows.
Section 2 following the recent experimental attempts to realize electron spin processor in silicon capable of quantum information processing introduces quantum model for excitation and coherent control of electron spin states via entangled proton trajectories. Exchange coupling is analyzed under quasiharmonic approximation of interaction potential taking into account the constraint of singular proton flux density. The theoretical model is further explained by Moliere's approximation of the Tomas Fermi interaction potential. This formalism comprises Liouville's theorem to give simple explanation for mapping procedures for proton beam transformation matrix in configuration and angular phase space. Section 3 explains numerical model and simulation parameters. Section 4 compares and discusses profiles of proton density distributions for transverse position plane (configuration space) and scattering exit-angle plane (angular space), gives the evolution of proton fluxes with various tilt angles and further illustrates comparative analysis of proton trajectories mapped in six dimensional phase space considering several $\Lambda$ and $\varphi$ variables in effective ion - crystal anharmonic potential. Mapping procedure for entangled proton trajectories is further obtained considering localization, selective excitation and unitary transformations of singlet/triplet spin states in quantum phase space.

# Results

Coherent manipulation and precise control of single electron spin rotation represents first step toward quantum information processing (QIP) [20, 32]. In order to achieve high level of precision of single electron spin unitary rotations we propose highly correlated spin chain of superfocused protons as a direct probe method for induction of local electron spin excitations in silicon. In this context propagation of the single spin excitation as a procedure for quantum state entanglement [33] can be intermediated via mixed quantum state between channeled protons (CP) spin system and induced coherent oscillations of electron spin system in silicon. In spin-lattice system, the condition of conservation of transverse energy when CP have equal probability to access any point of physical area corresponding to the channeling conditions, i.e., reaching the state of statistical equilibrium, has been modified by Barrett factor [34]. This constraint explains simultaneous existence of equilibrium particle distribution and population enhancement in different fractions of the phase space volume in process of ion transmission through media of sufficient small length. As a result, phase space distributions of CP in separate non-equipotential areas of the channel exhibit fractal characteristics over total phase space volume.

We have investigated the proton flux profile in scattering angle plane and transverse position plane. The boundary conditions of nonuniform density distributions are analyzed in case of small impact parameters along main $\langle 100 \rangle$ Si crystal axis. The obtained results show that enhancement effect of channeled protons flux bijectivelly corresponds to flux maxima in coordination space. In that sense we have analyzed degree of anisotropy including the anharmonic, higher order terms, $k^i$, $i \leq 4$, in effective continuum interaction potential. As a result, the channeled proton (CP) induced transition frequency $\omega$ between two electronic states includes higher order contributions

$$\hbar\omega = \hbar\omega - \frac{1}{2}|E_\perp(r,\varphi,\Lambda)|^2 (\alpha_2(\lambda,\varphi) - \alpha_1(\lambda,\varphi)) + O^{n \leq 4}. \tag{2.1}$$

$E_\perp$ represents the effective transverse energy of induced CP - electron system interaction. The corresponding Hamiltonian of rotating frame system [32] is

$$Hs = \Omega S_z + \omega_I I_z + A S_z I_z + B S_z I_z \tag{2.2}$$

$$\Omega = \omega_s - \omega, \ \omega_s = g\beta_e B_o / \hbar, \ \omega_I = -g_n \beta_n B_o / \hbar$$

$\Omega$ represents CP perturbed electron Zeeman frequency, $\beta_e$ and $\beta_n$ denote the Bohr and nuclear magneton respectively, $S_z$ and $I_z$ are electron and nuclear spin operators, $g$ and $g_n$ are electron and nuclear effective g factors, $\omega_s$ and $\omega_I$ denote electron and nuclear Zeeman frequency respectively (nuclear frequency for spin up polarization $\omega_{I\uparrow}$ = 11.99 and nuclear frequency for spin down polarization $\omega_{I\downarrow}$ = 36.35), $A$ and $B$ are secular and pseudosecular hyperfine couplings terms, respectively, $\hbar$ is reduced Planck constant, $B_o$ is static magnetic field along z-axis. Under static magnetic field the singlet $|S\rangle$ and polarized triplet $|T_{0,\pm}\rangle$ are degenerate, nearly independent. As a result, the quantum state of system $\rho$ in rotating frame corresponds to position of spin down polarized axis (bottom half of the Bloch sphere) as $\rho \otimes |0\rangle\langle 0| \otimes U$, likewise the polarized spin up axis position (top half of the Bloch sphere) denotes $\rho \otimes |1\rangle\langle 1| \otimes U$, where $U$ (Eq. (10)) couples additional degrees of freedom to initial quantum state, i.e. it represents the transformation matrix of mixed quantum state under CP polarization. In finite magnetic field, the CP perturbed electron Zeeman frequency for external field up to 1 MeV allows decorrelation of longitudinal Overhauser field $B_Z$ and shifts the level of singlet spin-down configuration from the ground into the excited $S|T_0\rangle$ and $S|T_+\rangle$ state. This coherent superposition of system energy levels (1, 0) and (0, 1) with triplet state (1, 1) is consistent with dipole-dipole mediated nuclear diffusion and leads to periodic superposition of spin states with precession period on a 1s time scale assuming that Overhauser fields $B = (\Delta B_X, \Delta B_Y, B_Z)$ are Gaussian distributed on long time scales. Thus, the external field close to 1 MeV is large enough to cause the strong spin dependency of tunneling effect. Large field produces strong asymmetry for spin up and spin down charge energy. It is important first to establish a non-zero external magnetic field $B$ so that each of nuclear spin principle axis orientations can be effectively optimized. This produces efficient coupling of longitudinal component of electron spins to quantized transverse component of nuclear spins. Thus, the quantized nuclear spin states are mediated via anisotropic part of the hyperfine interaction, i. e. the universal control of the nuclear spin state is achieved via unitarily transformed term, $BS_zI_z$ [35].

Namely, up to 1 MeV the external field induces coupling of nonparalell nuclear spin quantization axis to electron spin states and it allows the anisotropic pseudosecular term for universal control, otherwise the pseudosecular term is suppressed. Hence, adding a stronger external field to the quartic potential (Eq. (2.1)), alters the potential minima and changes confinement energies of orbital wave states which in turn induces a DOS transition from apsolute equilibria to saddle point in phase space.

Instead of applying the oscillating RF field to spatially resolve and manipulate spin resonance frequencies (or in order to measure response of the quantum dot by current flowing through the dot or by near quantum point), the induced transition can be generated upon CP excitation of the spin system. Thus, excited spin system displaces the center of the electron wavefunction along the oscillating superfocused CP field direction and change its potential depth. As a result the electron wavefunction frequency can be spatially distorted in order to coincides/shifts with applied CP field.

A single spin excitation is then polarized along z axis coinciding with proton beam alignment. In addition, the resultant mixed state conserves the total angular momentum of the exchange Hamiltonian along z axis:

$$[H_{ech}; Z] = 0, \quad Z = \sum_{j=1}^{n} \sigma_z^j. \tag{3}$$

This allows diagoanalization of the system Hamiltonian into subspaces of excited spins, i.e. the spin ensemble along $\sigma_z$ basis, corresponds to degenerate Z eigenvalues. The effective single spin read out [36] can be further realized by electrical detection of spin recharge events in tunneling proximity to a metal by adjustment of the Fermi level between two initially split electron eigenstates (corresponding to spin-up and spin-down orientation). Excitations of electron spin localized below Fermi threshold causes electron tunneling and leaving of initially occupied eigenstate. Discharged, empty spin state below Fermi level is further filled by an electron with oppositely oriented spin.

In present case, numerical solutions of entangled proton trajectories, for different reduced crystal thicknesses and tilt angles, correspond to short range correlated proton - lattice interaction potential in vicinity of $\langle 100 \rangle$ Si axis.

The interaction between the proton and the crystal's atoms includes elastic collisions, assuming classical, small-angles model of channeling [1, 28]. For the zero φ angles, the z-axis coincides with $\langle 100 \rangle$ Si crystallographic axis, while the atomic strings which define the channel cover the x and y axis. The initial proton velocity vector $v_0$ is collinear with the z axis.

We have modeled the system considering the Lindhard continuum approximation for axial channeling [1]. The crystal interaction potential comprises the continuum potentials

of separate atomic strings. Hence, we have included the thermal vibrations of the crystal's atoms:

$$U_i^{th}(x,y) = U_i(x,y) + \frac{\sigma_{th}^2}{2}\left[\partial_{xx} U_i(x,y) + \partial_{yy} U_i(x,y)\right], \tag{4}$$

where $U_i(x,y)$ represents the continuum potential of the $i$th atomic string, $xy$ are transverse components of the proton position, and $\sigma_{th}$ is the one-dimensional thermal vibration amplitude.

The specific electronic energy loss is determined by equation

$$-\frac{dE}{dz} = \frac{4\pi Z_1^2 e^4}{m_e v^2} n_e \ln \frac{2 m_e v^2}{\hbar \omega_e}, \tag{5}$$

where $v$ is the proton velocity, $m_e$ is the electron mass, $n_e = \dfrac{\Delta U^{th}}{4\pi}$ is the density of the crystal's electrons averaged along the z axis and $\Delta \equiv \partial_{xx} + \partial_{yy}$.

The angular frequency of the electron oscillation induced by the channeled proton is

$$\omega_e = \left(\frac{4\pi e^2 n_e}{m_e}\right)^{\frac{1}{2}}. \tag{6}$$

The mean-square angular deviation of the proton scattering angle caused by its collision with the electrons is included as

$$\frac{d\Omega}{dz} = \frac{m_e}{2 m_p E}\left(-\frac{dE}{dz}\right). \tag{7}$$

In the above equation $m_p$ denotes the proton mass and $E$ is the proton energy.

Further calculations take into account the proton beam divergence before its interaction with the crystal [24, 25].

The Monte Carlo simulation method has been used for parameterization of entangled proton trajectories. Obtained numerical solutions of channeled protons equations of motion correspond to their angular and spatial distributions. The phase space density, according to the Liuoville´s theorem, cannot be changed in conservative system, but one can manipulate with the form and position of the phase space elements. We can use the phase space transformations to improve the channeling efficiency.

Discrete map of quantum states of channeled proton trajectories, their point transformation in the spatial (transverse) and angular phase space are presented in the following vector basis

$$\begin{pmatrix} x \\ y \\ x' \\ y' \\ \Lambda \\ \varphi \end{pmatrix} \rightarrow \begin{pmatrix} x \\ y \\ \theta_x \\ \theta_y \\ \Lambda \\ \varphi \end{pmatrix} \qquad (8)$$

Here $\Lambda$ and $\varphi$ denote the crystal reduced length and tilt angle of CP beam.

Although the phase space is six dimensional we consider four subspaces of the transverse and angular phase space. Correspondingly, the mapping of beam parameters and quantum discretization of entangled states is conformally rescaled through phase space transfer matrix $\mathbf{M}$ (4x4): $\mathbf{M} \rightarrow J\widetilde{\mathbf{M}}$

In that sense we consider local symplectic condition for the realizable transfer matrix of the Hamiltonian system

$$\widetilde{\mathbf{M}} J \mathbf{M} = \mathbf{M} J \widetilde{\mathbf{M}} = J, \qquad J = \begin{pmatrix} J_{2D} & 0 \\ 0 & J_{2D} \end{pmatrix}, \quad J_{2D} = \begin{pmatrix} 0 & 1 \\ -1 & 0 \end{pmatrix} \qquad (9.1)$$

Tilde sign denotes the transpose operation over transfer matrices and $\mathbf{J}_{2D}$ refers to unit symplectic matrix in 2-d phase space volume.

According to the Liuoville's theorem the conservation of the phase space volume results from the statement: $\det \mathbf{M} = 1$, following the equation (9.1).

Complete characterization of the phase space volume is achieved over the second order moments of transfer beam matrices:

$$\Sigma(X) = \langle X \widetilde{X} \rangle = \begin{pmatrix} \langle x^2 \rangle & \langle x'x \rangle & \cdots & \langle \varphi x \rangle \\ \langle xx' \rangle & \ddots & & \vdots \\ \vdots & & \ddots & \\ \langle x\varphi \rangle & \cdots & \cdots & \langle \varphi^2 \rangle \end{pmatrix}, \quad \Sigma(\vartheta_X) = \langle \vartheta_X \widetilde{\theta_X} \rangle = \begin{pmatrix} \langle x^2 \rangle & \langle \theta_x x \rangle & \cdots & \langle \varphi x \rangle \\ \langle x\theta_x \rangle & \ddots & & \vdots \\ \vdots & & \ddots & \\ \langle \theta_x \varphi \rangle & \cdots & \cdots & \langle \varphi^2 \rangle \end{pmatrix}, \qquad (9.2)$$

$$\Sigma(Y) = \langle Y \widetilde{Y} \rangle = \begin{pmatrix} \langle y^2 \rangle & \langle y'y \rangle & \cdots & \langle \varphi y \rangle \\ \langle yy' \rangle & \ddots & & \vdots \\ \vdots & & \ddots & \\ \langle y\varphi \rangle & \cdots & \cdots & \langle \varphi^2 \rangle \end{pmatrix}, \quad \Sigma(\vartheta_y) = \langle \vartheta_y \widetilde{\theta_y} \rangle = \begin{pmatrix} \langle y^2 \rangle & \langle \theta_y y \rangle & \cdots & \langle \varphi y \rangle \\ \langle y\theta_y \rangle & \ddots & & \vdots \\ \vdots & & \ddots & \\ \langle \theta_y \varphi \rangle & \cdots & \cdots & \langle \varphi^2 \rangle \end{pmatrix}. \qquad (9.3)$$

The beam matrix transform as

$$\Sigma \rightarrow M \Sigma \widetilde{M} \quad (9.4)$$

$$I' = -1/2 Tr(\Sigma J \Sigma J), \sigma_4 = \det(\Sigma) \quad (9.5)$$

where matrix trace, $Tr$ and value $\sigma_4$, i.e. the phase space volume occupied with the proton beam, determines the two invariants of the transfer beam matrix.

We reduce system dimensionality by decoupling the 4-d phase space on 2-d: configurational $(x-y)$ and angular, $(\theta_x - \theta_y)$ phase space. The transformation matrix describing the mixed quantum ensamble then couple a digonal $S_X$ basis of electron spin system to diagonal $S_Z$ basis [37] of fully polarized superfocused CP beam and forms a nonothonormal basis $U = \sum_r |\Sigma^r\rangle\langle a^r|$. (10)

Basis $|\Sigma^r\rangle = \sum_{nm} \alpha_{nm} e^{-i\varepsilon_{nm}t} |nm\rangle$ denotes the stationary, $|nm\rangle$ quantum state of electron system with amplitude $\alpha_{nm}$, energy $\varepsilon_{nm}$ and quantum numbers n, m.

Under CP interaction $i\rho(\alpha_{nm}) = \sum_{n'm'} \langle nm|U|n'm'\rangle e^{i(\varepsilon_{nm}-\varepsilon_{n'm'})t} \alpha_{n'm'}$ denote energy splitting between ground and excited electron states, $|n'm'\rangle$, of target and projectile.

In order to determine the matrix elements of CP - lattice confinement potential for singlet and triplet functions we use two single electron eigenstates denoted by spatial electron-waive functions $|X\rangle$ and $|X'\rangle$ as

$$|S\rangle = |\uparrow\downarrow\rangle - |\downarrow\uparrow\rangle \otimes \begin{pmatrix} 1/\sqrt{2}(|XX'\rangle) \\ 1/2(|XX'\rangle + |X'X\rangle) \end{pmatrix}, \quad |T_{0,\pm 1}\rangle = \begin{pmatrix} 1/2(|XX'\rangle - |X'X\rangle) \otimes |\uparrow\downarrow\rangle + |\downarrow\uparrow\rangle \\ 1/\sqrt{2}(|XX'\rangle - |X'X\rangle) \otimes |\uparrow\uparrow\rangle \\ 1/\sqrt{2}(|XX'\rangle - |X'X\rangle) \otimes |\downarrow\downarrow\rangle \end{pmatrix}$$

The energy splitting between triplet, $|T_0\rangle$ and ground singlet state, $|S\rangle$ is denoted by exchange interaction, $J = \langle T_0|i\rho(\alpha_{nm})|T_0\rangle - \langle S|i\rho(\alpha_{nm})|S\rangle$. (11)

The Hamiltonian is further diagonalized in singlet and triplet subspaces. In order to overcome high level truncation of the basis, where linear combinations of two electron states tends to infinity, we use constraint that singlet state refers to ground state according to Lieb Mattis theorem [38] in zero magnetic field.

Applying the inhomogeneous CP field along main crystal axis, i.e., involving x and y phase space components of tilted CP beam, if the energy difference of triplet and singlet electron states is close, they became strongly mixed. The triplet $|T_0\rangle$ can evolve into the singlet state $|S\rangle$ as

$$1/2(|XX'\rangle-|X'X\rangle)\otimes|\uparrow\downarrow\rangle+|\downarrow\uparrow\rangle \rightarrow |\uparrow\downarrow\rangle-|\downarrow\uparrow\rangle\otimes 1/2(|XX'\rangle+|X'X\rangle);$$

Likewise, $|T_+\rangle$ and $|T_-\rangle$ evolve into singlet state. As explained in the main text the mechanism of spin excitation and energy separation scheme, (illustrated in figure 5), between the ground singlet state and the polarized triplet state is controlled by a combination of a CP initial energy E ($\varphi$, Λ) and tilt angle $\varphi$. Upon the excitation energy is applied to the quantum dot inside the Bohr radius, it is shown that spin system energy cost for adding an extra electron starts from state S (0, 1), as indicated by dotted black line, where (n, m) J(ε) and E$_{ST}$ denote the charge state with n and m electrons, exchange and splitting energy, respectively. The energy cost for reaching (1, 1) is (nearly) independent of the spin configuration.

However, the energy cost for forming a singlet state S (0, 2) is much lower than that for forming a triplet state (not shown in the diagram). This difference can be further exploited for spin initialization and detection.

## Discussion

**Figure 1 (a (1, 2), b (1, 2), c (1, 2))** gives 3-d representation of channeled protons contour plots for 92 nm $\langle 100 \rangle$ Si nanocrystal, Λ = 0.5 [24, 25], for tilt angles: $\varphi = 0.05\psi_c$, $\varphi = 0.15\psi_c$ and $\varphi = 0.20\psi_c$, $\varphi$ is the angle of external field relative to symmetry axis of the spin transformation tensor. The external field of 1 MeV is chosen to match the limits of Bohr radius with initial CP peak separation at 20% of the critical angle (relative to tensor principal axis). It allows generation of the final mixed quantum state, controlled by the pseudosecular term $B = 3D\cos(\varphi)\sin(\varphi)$, i.e. it allows efficient dipolar coupling, $D$ between electron and nuclear spin states. The spacing between separate peaks about longitudinal z-direction of the confinement field is calculated via $\frac{\pi^2\hbar^2(n+1/2)}{(L^2 m_e)}$, L = 92 nm, for n-discretized 2-d potential: $E = E_n\left(n_{(X,\theta_x)}, n_{(Y,\theta_Y)}\right)_Z$. The obtained results show density of states (DOS) evolution in phase space for angular and transverse position profiles. The analysis for tilts $\varphi \leq 0.05\psi_c$, figure a (1, 2), shows the area of maximal enhancement in ion flux density [27] in both phase planes. The maximal confinement field is governed by exchange coupling energy, $J(\varepsilon)$. Here $J(\varepsilon)$ represents the function of energy difference, $\varepsilon$ for discretized 2-d potential. Figure b (1, 2) shows that incident tilt angles above 15 % of the critical angle for channeling, consequent faster amplitude and phase attenuation for angular density profile. This effect induces further splitting of the channeling pattern.

Relative change of 5% for crystal tilts leads to strong yield redistribution in angular distribution profiles and it mostly affects phase profile central parts. The analysis in configuration space for tilts: $\varphi = 0.05\psi_c$ and $\varphi = 0.15\psi_c$ shows strongly picked circular cross section. Only a slight variation in pattern sharpness can be seen in density profile edges. DOS analysis for $0.20\psi_c$ reveals further effects of strong perturbation, as presented in figure 1(c (1, 2)). The angular phase space profile, 1(c (1)) shows non-homogenous transition in charge state density and splitting of channeled proton distribution pattern. The characteristic splitting shows two pronounced maxima on the $\theta_x$ - axis followed by few nonsymmetrical peaks as lateral satellites. Their spatial positions and amplitudes are correlated via CP mediated Zeeman interaction by $g\beta_e B/2$ term. This non-secular term shifts the energy levels of singlet spin states and splits DOS peaks (shift affects the Fermi level for electron spin-down and spin-up configuration). Consequently, spin DOS structure is uniquely described via two mixed quantum states.

**Figure 2 (a, b)** shows the central position's of angular and spatial density distributions in phase space. Maximal amplitudes of tilt angles equals $\varphi = 0.05\psi_c$, $0.10\psi_c$ and $0.15\psi_c$. In figure 2(a) designated plots correspond to reduced crystal thickness in range of 0.00 - 300.0., for $L$ = 1.69 µm and 0.00 - 0.300, for $L$ = 99.2 nm, respectively. These dependencies determine the focusing region, i.e. specify the proton beam full with at half maximum (FWHM).

A comparative analysis for the same values of $\varphi$, $\Lambda$ and amplitude maxima in configuration plane is presented in figure 2(b), it determines the length and phase space transformation bond between scattering angle plane and mapped transverse position (configuration) plane.

**Figure 3** shows that yield dependence of harmonic confinement potential (governed by first two terms in Eq. 2.1) becomes zero for tilts over $0.50\psi_c$ in transverse and angular phase plane. The normalization and boundary condition are restricted to effective Bohr radius: $a^* = \hbar^2 k / m_e$. Changing tilts while keeping fixed thickness parameter to $L = 99.2$ nm gives the non-monotonic dependence to exchange coupling energy $J(\varepsilon)$ as a function of quantum displacement from harmonic oscillator stability point. It goes to zero asimptotically and indicates the complete separation of quantum states and inexistence of singlet-triplet transitions at higher tilts due to small orbitals overlap (Eq. 16, 17).

**Figure 4** illustrates the localization of quantum spin waves corresponding to uncertainty principle. The CP superimposed electron spin states (spin wave probability density functions) are positioned inside the Bohr radius. Electron probability densities produce maxima over each nuclear position.

The quantum proton trajectory evolution with various tilt angles is calculated for $\Lambda = 0.25$ in configuration space, i.e. $\Lambda = 0.50$ in (mapped) angular space. We analyze eight characteristic tilt shifts: $\varphi = 0.00, 0.05\psi_c, 0.10\psi_c, 0.15\psi_c, 0.20\psi_c, 0.25\psi_c, 0.35\psi_c, 0.50\psi_c$. Inside the Bohr radius at distance $x = \pm d$ around the peak centers, the confinement field is parabolic so that ground state of mixed wave functions coincides with harmonic oscillator state.

Observed amplitude dependences of proton yield for tilts $\geq 0.10\psi_c$ are attributed to stronger interaction influences of higher anharmonic terms in Eq. (2.1). This is more pronounced for the spatial CP distribution.

Amplitude decreasing and changes of peaks FWHM (positions and spatial symmetry) are indicators for the state of strong system perturbation, i.e. due to the effect of quartic anharmonic terms in the exchange interaction, Eq. 2.1.

The observed modulation of DOS states for tilts $\geq 0.20\psi_c$ is disregarded, i.e. the main contribution to the superfocusing effect comes from the crystal tilts below 20% of the critical angle for channeling.

The analysis of the asymptotic behavior of the channeled protons axial yield for proton distributions: $\varphi = 0.00\psi_c$, $\Lambda = 0.5$, (when FWHM of generated focused area converges to zero making the sub-nanometer spatial resolution possible) has shown that the only case when the angular yield is singular corresponds to the zero-degree focusing effect. It is shown that an increase in the crystal tilt angle value of 15% of the critical angle for channeling facilitates the suppression of the zero-degree focusing effect. Former leads to a significant change in amplitude and width of the angular channeled protons profile, causing the splitting on two lateral non-uniform circular patterns with maxima located along $\theta_x$-axis which corresponds to smaller lateral peaks.

This behavior is confirmed for energy range from several eV up to several MeV which can be employed for PIXE analysis. To facilitate and control close encounter collision processes in order to induce nuclear reactions one can also use this method as intermediate process for nuclear collision cascade. However for the proton beam energies above 100 MeV the FWHM of the channeled proton peaks in density profiles is highly narrower $\leq 5\,\text{pm}$ less then effective Bohr radius due to enhanced orbitals overlap of superfocusing effect governed by higher degree of spatial confinement.

**Figure 5** represents the scheme of energy splitting and exchange coupling energy between singlet $|S\rangle$ state and $|T_-\rangle$ triplet-polarized state localized inside the Bohr radius. Upon modulation the CP field, unitary spin rotations are performed around two non-commuting axes: $\theta, z$. Before manipulation, discretized proton spin states mediated through Zeeman interaction include sublevels $|01\rangle$ and $|10\rangle$. The populations of quantum states are distributed according to electron spin polarization at thermal equilibrium: in electron manifold the nuclear quantization axes $-\frac{B}{2}I_{(X,Y,\theta_X,\theta_Y)} + \left(\omega_n - \frac{A}{2}\right)I_Z$ correspond to electron spin in $|\downarrow\rangle$ state, whereas the nuclear axes, $+\frac{B}{2}I_{((X,Y,\theta_X,\theta_Y))} + \left(\omega_n + \frac{A}{2}\right)I_Z$ apply to electron spin in $|\uparrow\rangle$ state. We then impose a pulse sequence of $\pi/n$ tilts relative to $\vartheta$ axes on Bloch sphere. This lifts the system energy close to triplet state where exchange $J(\varepsilon)$ is large. It triggers the coherent transition of proton eigenstates $|01\rangle \rightarrow |\uparrow\rangle$, $|10\rangle \rightarrow |\downarrow\rangle$ and forms the final mixed quantum entangled state containing the superposition of proton - electron eigenstates $|\downarrow\uparrow\rangle, |\uparrow\downarrow\rangle$.

To provide a transition to triplet state $|\downarrow\downarrow\rangle$ upon initialization the system is rotated by $\pi$ pulse about z axes of Bloch sphere through the angle $\varphi = J(\varepsilon)/\hbar$, where $J(\varepsilon)$ denotes the exchange coupling as a function of energy difference, $\varepsilon$ between the levels. Presented energy diagram shows that former sequence corresponds to the initial exchange splitting $E_{ST}$ further dominated by CP induced $J(\varepsilon)$ mixing between energy levels detuned by $\varphi$. While increasing the confinement energy $E_{(\varphi,\lambda)}$, the (1, 1) triplet state hybridizes and produces a tunneling effect so that different superposition states can be realized.

Dependency of the field caused by lifted degeneracy of triplet state further decreases the separation between energy levels, while the exchange coupling increases the Gaussian of orbitals overlap (Eq. 16).

**Figure 6 (a, b, c)** shows the CP simulation patterns in transverse position plane for fixed value of $\Lambda = 0.175$. The proton trajectory, shifts with tilt angle along y = 0 axis and spreads from intersection area of x-y plane into cusped elongated deltoidal pattern, figure 6 (a).

The shifts are governed by atomic strings repulsive potential. Even a small change of tilt angle causes a strong system perturbation evident in figure 6(b) and therefore activates the higher power terms in the ion-atom interaction potential. Former influences regularity of proton trajectories and leads to a gradual reduction of DOS in central area of channel. This causes the non-uniform flux redistribution, now filled with gaps, as it can be seen from figure 6 (c). Hence, it affects the continuous conservation of distribution functions in phase space volume [28, 29]. Consequently, the axially channeled protons cannot encounter the state of statistical equilibrium. That effect has been resolved in scope of KAM theory [49], when the classical integrability of the Hamiltonian's is broken by sufficiently small perturbation, the system nevertheless retains its dynamics in the form of periodic oscillation moving on invariant phase space profile. Although these invariants of phase space have form of intricate fractal structure in vicinity of $\langle 100 \rangle$ axis, they still cover a large portion of phase space. In that sense the reduced crystal thickness can be fully discretized by performing the power low expansion of random points described for interval $[n_i, n_i+n]$ between two nearest neighboring fractal points $\Lambda(n) = Ln^\alpha$, $n \to \infty$, where fractal dimension for $\alpha < 0$ draws logarithmic singularity for proton density distribution.

## Methods

Simulation model considers cubic unit cell representation of the isotopically pure $^{28}$Si nanocrystal. It includes atomic strings on three nearest square coordination lines of the $\langle 100 \rangle$ axial channel [24, 25, 46, 47]. According to diamond lattice symmetry, the orthogonal mesh is projected across the channel, mapping two layers of 2x2 triangular areas of $\langle 100 \rangle$ unit cell.

The proton trajectories are generated from the sequences of binary collisions via the Monte Carlo simulation method using the screened Moliere's interaction potential. The crystal is tilted in angular space along the axis $\theta_x = 0$, $(x = 0)$ where the value of the tilt angle ranges up to 20% of the critical angle for channeling, $\psi_c = \left[ 2Z_1 Z_2 e^2 / (dE) \right]^{1/2} = 6.09$ mrad. Numerical calculations consider the continuum model in the impulse approximation.

The motion of ions in the continuum model is determined by the Hamiltonian

$$H = (1/2) m \left( p_\perp^2 + U(r) \right) = E \left( \psi_x^2 + \psi_y^2 \right) + U(r), \tag{12}$$

$$E_\perp = E\psi^2 + U(r), \tag{13}$$

where $\psi_x$ and $\psi_y$ denote $x$ and $y$ projection of scattering small angle with respect to the $\langle 100 \rangle$ axis. The systems ion-atom interaction potential is obtained by integration of the Moliere's approximation of the Tomas Fermi interaction potential [39].

$$U_i(r) = \frac{2Z_1 Z_2 e^2}{d} \sum_{i=1}^{3} \alpha_i K_0 \left( \beta_i \frac{r}{a} \right). \tag{14}$$

$Z_1$ and $Z_2$ denote the atomic numbers of the proton and the atom, respectively, $e$ is the electron charge, $d$ measures quantum displacement of single particle wave function relative to harmonic oscillator central position in ground state, $r$ is distance between the proton and separate atomic strings, $a_0$ is the Bohr radius and $a = \left[ \frac{9\pi^2}{128 Z_2} \right]^{\frac{1}{3}} a_0$ gives the atom screening radius; $K_0$ denotes the zero order modified Bessel function of the second kind, with the fitting parameters: $(\alpha_i) = (0.35, 0.55, 0.10)$, $(\beta_i) = (0.30, 1.20, 6.00)$.

Correspondingly, the potential between two $i, j$ sites is $\Phi_{i,j} = U_i(r) + B/r_{ij}^n$. (15)

$n$ is Born exponent. Coefficients $B$ and $n$ are experimental fitting parameters determined from ion compressibility measurements [40], likewise, the exponential repulsion between the overlapping electron orbitals within the channel is described by $B \exp\left( -\rho / r_{i,j} \right)$.

The measure of the orbitals overlap corresponds to $l = \exp\left( d^2 / a^2 \left( (1/b) - 2b \right) \right)$ (16)

while $b = \sqrt{(1 + \omega_s / \Omega)} \, d$ and $\varepsilon = (d/a) f \hbar \omega_e$ denote degree of confinement field and dicretized energy, respectively. (17)

Eq. (16) includes the variation of charge density of the overlapping area due to different valence electron contribution to interaction: lattice - induced potential (across the channel) [41, 42]. The overlap $l = exp(d^2/a^2)$ stands only for zero external field. The one electron energies [43] for neutral Si: $(1s)^2$ $(2s)^2$ $(2p)^6$ $(3s)^2$ $(3p)^2$ are calculated in Hartree unit 1 Hartree = 2 Rydberg = 27.210 eV [44] as 1s = 67.02, 2s = 5.5435, 2p = 3.977, 3s = 0.49875, 3p = 0.2401. The electron affinity for Si crystal is 4.018 eV and 1.385 eV for Si atom [43].

In the rotating frame, the reduced form of protons equation of motion, considering small angle approximation in transverse position plane [39], is

$$\frac{\partial x}{\partial z} = \varphi_x, \quad \frac{\partial y}{\partial z} = \varphi_y, \quad \frac{\partial \varphi_x}{\partial z} = -\frac{1}{2E_\perp}\frac{\partial U(x,y)}{\partial x}, \quad \frac{\partial \varphi_y}{\partial z} = -\frac{1}{2E_\perp}\frac{\partial U(x,y)}{\partial y}. \tag{18}$$

$\varphi_x$ and $\varphi_y$ represent the x-y component of the proton scattering angle. The channeled proton distributions are mapped in configuration space and angular space in two steps: to transverse position phase plane, $x'$ - $y'$ and to scattering angle phase plane, $\theta_x$ - $\theta_y$ [24, 25] in accordance with the chosen value of reduced crystal thickness, $\Lambda$ and the tilt angle, $\varphi$.

The phase space transformations are determined via Jacobean:

$$J_\theta \equiv \frac{\partial \theta_x}{\partial x} \cdot \frac{\partial \theta_y}{\partial y} - \frac{\partial \theta_y}{\partial x} \cdot \frac{\partial \theta_x}{\partial y}, \quad J_{\theta_{x,y}} = J(x,y,\varphi,\Lambda). \tag{19}$$

Eq. (19) comprises the proton trajectory components: $\theta_x(x,y,\varphi,\Lambda)$ and $\theta_y(x,y,\varphi,\Lambda)$. It establishes a bond transformation between differential transmission cross section, $\sigma = 1/|J|$ and phase space manifolds in configuration and angular plane.

The one-dimensional thermal vibration amplitude of the crystal's atoms is 0.0074 nm [24, 25, 29, 45, 46]. The average frequency of transverse motion of protons moving close to the channel axis is equal to $5.94 \times 10^{13}$ Hz. It is determined from the second order terms of the Taylor expansion of the crystal continuum potential in vicinity of the channel axis [48, 49]

$$U(x,y) = \frac{2Z_1 Z_2 e^2}{d} \sum_{i=1}^{3} \sum_{j=1}^{M} \alpha_i K_0\left(\beta_i \frac{\rho_c}{a}\right), \tag{20}$$

where $\rho_c = \sqrt{(x-x_i)^2 + (y-y_i)^2}$,

$$\frac{\partial^2 U(x,y)}{\partial x^2} = \frac{2Z_1 Z_2 e^2}{d} \sum_{i=1}^{3} \sum_{j=1}^{M} \alpha_i \frac{\beta_i}{a} \left( \begin{array}{c} K_0\left(\frac{\beta_i \rho_c}{a}\right)\left(\frac{\beta_i}{a}\right)\frac{(x-x_j)^2}{\rho_c^2} + \\ + \frac{2(x-x_j)^2 - \rho_c^2}{\rho_c^3} K_1\left(\frac{\beta_i \rho_c}{a}\right) \end{array} \right), \tag{20.1}$$

$$\frac{\partial^2 U(x,y)}{\partial y^2} = \frac{2Z_1 Z_2 e^2}{d} \sum_{i=1}^{3} \sum_{j=1}^{M} \alpha_i \frac{\beta_i}{a} \left( \begin{array}{c} K_0\left(\frac{\beta_i \rho_c}{a}\right)\left(\frac{\beta_i}{a}\right)\frac{(y-y_j)^2}{\rho_c^2} + \\ + \frac{2(y-y_j)^2 - \rho_c^2}{\rho_c^3} K_1\left(\frac{\beta_i \rho_c}{a}\right) \end{array} \right), \tag{20.2}$$

$$\Delta(U(x,y)) = \frac{2Z_1 Z_2 e^2}{d} \sum_{i=1}^{3} \sum_{j=1}^{M} \alpha_i \frac{\beta_i}{a} \left( \left( K_0\left(\beta_i \frac{\rho_c}{a}\right) \right) \left( \frac{\beta_i}{\rho_c a} \right) \right). \tag{21}$$

$K_1$ denotes the first order modified Bessel function of the second kind, $d$ and $M$ represent distance from atomic strings and their number, respectively.

The components of the proton scattering angle, $\varphi_x = \frac{v_x}{v_0}$ and $\varphi_y = \frac{v_y}{v_0}$, are solved numerically using the implicit Runge Kutta method of the fourth order [48].

The components of the proton impact parameter are obtained randomly from the uniform distributions inside the channel. The transverse components of the final proton velocity, $v_x$ and $v_y$ are presented within the Gaussian distribution of probability that the quantum spin state is recognized correctly, according to standard deviation $\Omega_{bx} = \Omega_{by} = \frac{\Omega_b}{\sqrt{2}}$, where $\Omega_b$ denotes the CP divergence. Since the channeled protons angular distributions can be easily measured, they are used to reconstruct the quantum information regarding the protons distribution in transverse phase space.

In order to quantify the read out fidelity the information from entangled quantum trajectories is sampled from $\theta_x$-$\theta_y$ phase plane from 550,000 shots datasets. The initial number of protons correspond to quantum trajectories spin states obtained from $5\times10^7$ traces.

To summarize our calculations and simulation results demonstrate a hybrid proton-electron quantum interface for multipartite entanglement under constraint metric of uncertainty principle. We established the correlation between electronic spin states and off-diagonal hyperpolarized nuclear spin states under CP induced field. We used axial configuration of Si$\langle 100\rangle$ channel to initialize and control each electron spin state via superimposed proton spin chain. Utilizing a dynamically decoupled sequence we have obtained the universal quantum control and controllable coupling between singlet and triplet-polarized spin states.

By calculating the electron spin and CP field eigenstates via full density matrix we established the proof of non-orthogonal mixed quantum state. Upon hyperpolarization sequence, the increased sensitivity of nuclear spin subspaces dependence to electron spin states reduces the linear spin entropy and leads to maximized entanglement of mixed states in density matrix. We have shown that stability dependence of nuclear field results from anisotropic term of the hyperfine coupling, here regarded as a tunable parameter for unitary spin control. It can be chosen to enhance the feasibility of producing entangled mixed states. A resultant mixed quantum state that we demonstrated in S-T systems represents important step toward realization of scalable architecture for quantum information processing. Complementary, a scalable network of entangled electron-nuclear states would form a basis for a cluster state of quantum processors integrated in silicon. In addition, generation of entanglement process comprising the network of such correlated spin states would enhance the quantum error correction beyond any separable state and extend the precision in quantum metrology. That would allow implementation of quantum error correcting techniques (QEC codes) directly to perfectly entangled mixed states and direct protection of quantum states from interaction with environment without prior entanglement purification protocols (EPP). In that context, the off-diagonal electron-nuclear eigenstates as mixed quantum states are not longer invariant under unitary spin operations and represent observables in density matrix.

Finally, the controllable addressing of single spins in quantum networks, the individual control of unitary spin precessions (electron-nuclear spin phase rotations) in combination with local g-factor engineering would provide a scheme for deposition of multipartite entangled states and manipulation of quantum memory and quantum key distribution (QKD) based on transmission of gaussian-modulated individual coherent states.

Another possibility for further exploration points toward active control of the channeled proton beam properties in the superfocusing effect, revealing the important role of mutual contribution of the harmonic and anharmonic terms. This emphasizes the importance in careful selection regarding the appropriate combination of the crystal tilt angle value with crystal thickness in order to gain high spatial resolution and localization accuracy. As a result the implementation of such nano-scale precision scanning method could produce a detailed map of discrete inter-atom positions, and create a highly resolved image, built-up through a process of the proton beam focusing.

This behavior is confirmed for energy range from several eV up to several MeV which can be employed for PIXE analysis. To facilitate and control close encounter collision processes in order to induce nuclear reactions one can also use this method as intermediate process for nuclear collision cascade. However for the proton beam energies above 100 MeV the FWHM of the channeled proton peaks in density profiles is highly narrower $\leq 5\,\mathrm{pm}$ less then effective Bohr radius due to enhanced orbitals overlap of superfocusing effect governed by higher degree of spatial confinement.

## Acknowledgments

This work is developed considering classical theoretical framework of dr. N. Nešković and dr. S. Petrović and helpful suggestions of dr. M. Rajković.

[27] For transverse energy $E_\perp < E\varphi$ and the beam incident angle $\varphi \ll \psi_c$, where $\varphi$ and $\psi_c$ denote the critical angle for channeling and effective ion atom potential, respectively. The effective potential area $\gamma\left(A_{l_0}(E_\perp)\right)$ which corresponds to maximal enhancement in ion flux density includes strictly harmonic terms under continuum approximation

$$\gamma\left(A_{l_0}\right) \approx \int_{A_{\min}}^{A_{\max}} \frac{dA}{A_{l_0} - \left(\pi E\varphi^2\right)/k} \approx \ln\left|\frac{A_0}{A_{l_0} - \left(\pi E\varphi^2\right)/k}\right|, \qquad (1.31)$$

where $A_{l_0}$ denotes the equipotential surface closed by field contour in the central part of the axial channel. The corresponding integration boundaries are:

$$A_{\min} = A_{l_0} - \frac{\pi E\varphi^2}{k} \quad \text{and} \quad A_{\max} = A_0 - A_i - \frac{\pi E\varphi^2}{k}.$$

$A_0 = \pi S_0^2 \alpha^{-1}$ and $A_i = \pi \rho_{cm}^2 \alpha^{-1}$ denote demarcation line of the channel cross-section area. The central part of the axial channel is then represented by the annulus of inner radius $S_0 = (\pi N d)^{-1} = \sqrt{\left(\frac{d}{2}\right) + \alpha \rho_c^2}$, where $d$, $\rho_c$ and $\alpha$ represent the mean spacing between the atomic rows, ion impact parameter and the ratio of total axial channels number versus number of atomic rows forming the channel, respectively.

Accordingly to equations (1.31), when the ion beam incident angle is close to zero $\varphi \cong 0$, the anisotropy for central part of axial channel is induced only by harmonic component of the interaction potential. This implies that first equipotential circle represents dominant effective potential area for ion flux density denoted as $\gamma\left(A_{l_0}(E_\perp)\right)$. Hence, the area of maximal enhancement in ion flux density is confined to central equipotential curve of axial channel $\gamma\left(A_{l_0}\right) \approx \ln\left|\dfrac{A_0 k}{\pi E \varphi^2}\right|$ and further converges to zero as $A_{l_0} \to 0$ if the incident angle, i.e. the tilt angle of the beam, corresponds to condition $\varphi = \sqrt{\left(A_{l_0} k / \pi E\right)}$. The results obtained for MeV proton beam energies show the nonequlibrium density of states across central part of the channel as nonuniform flux redistribution. This reveals the strong effect of anharmonic components in effective continuum interaction potential even in vicinity of low index crystal axis for $\langle 100 \rangle$ Si.

# Figure 1

## a (1, 2)

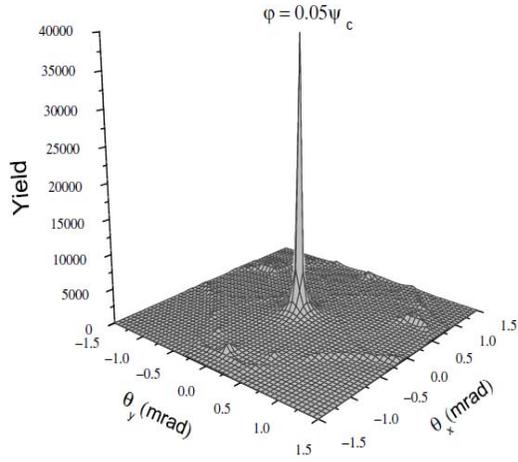
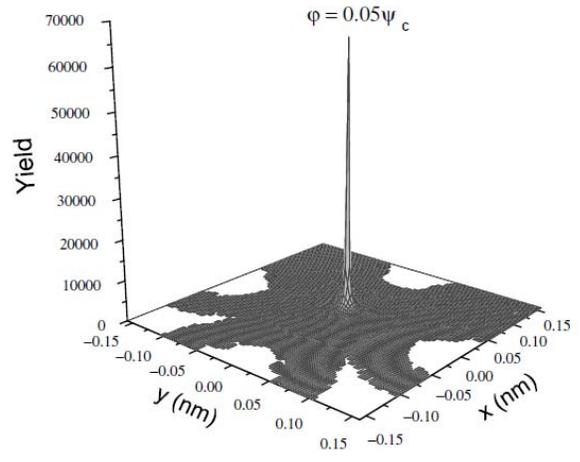

## b (1,2)

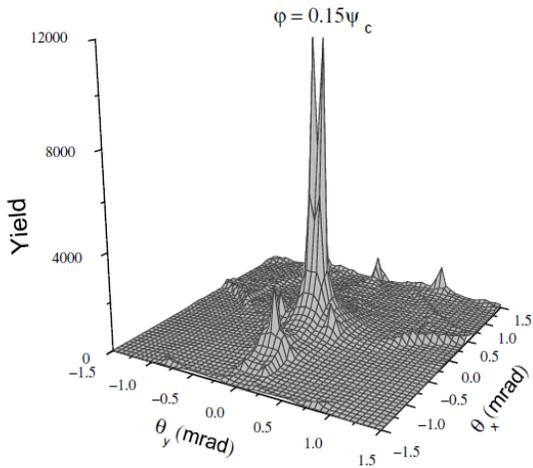
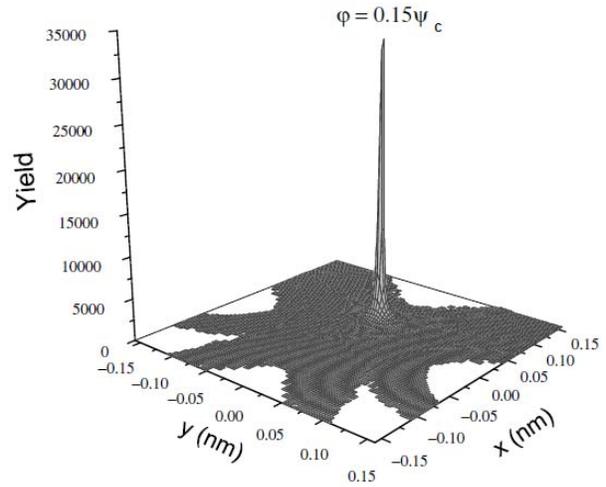

## c (1,2)

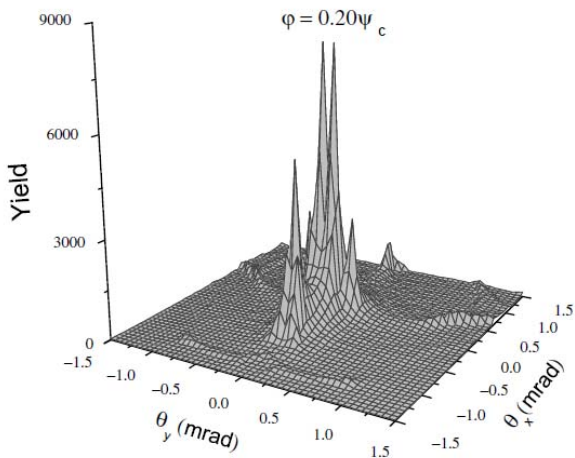
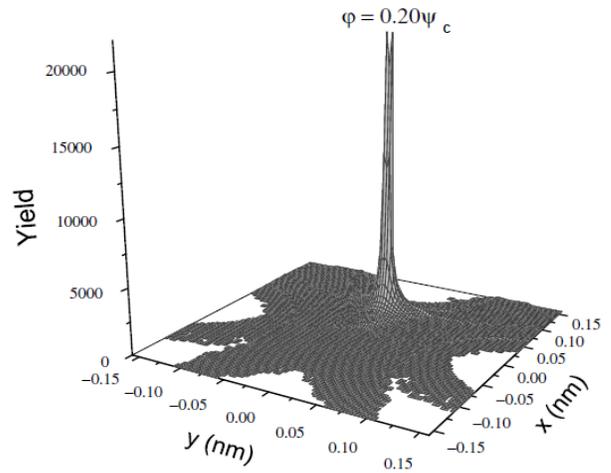

**Figure 2**

a)

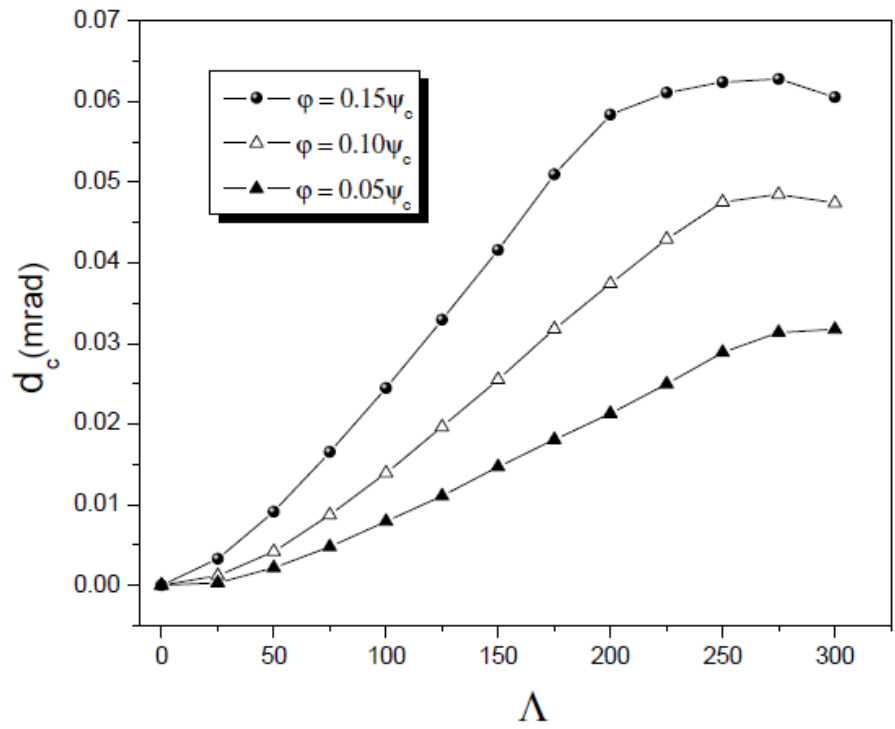

b)

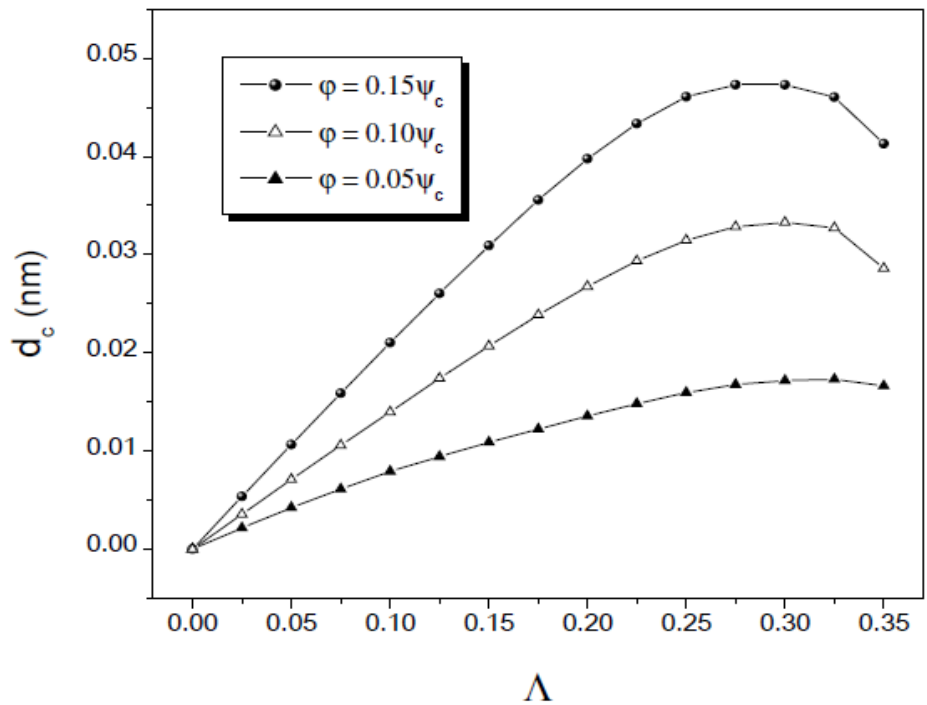

**Figure 3**

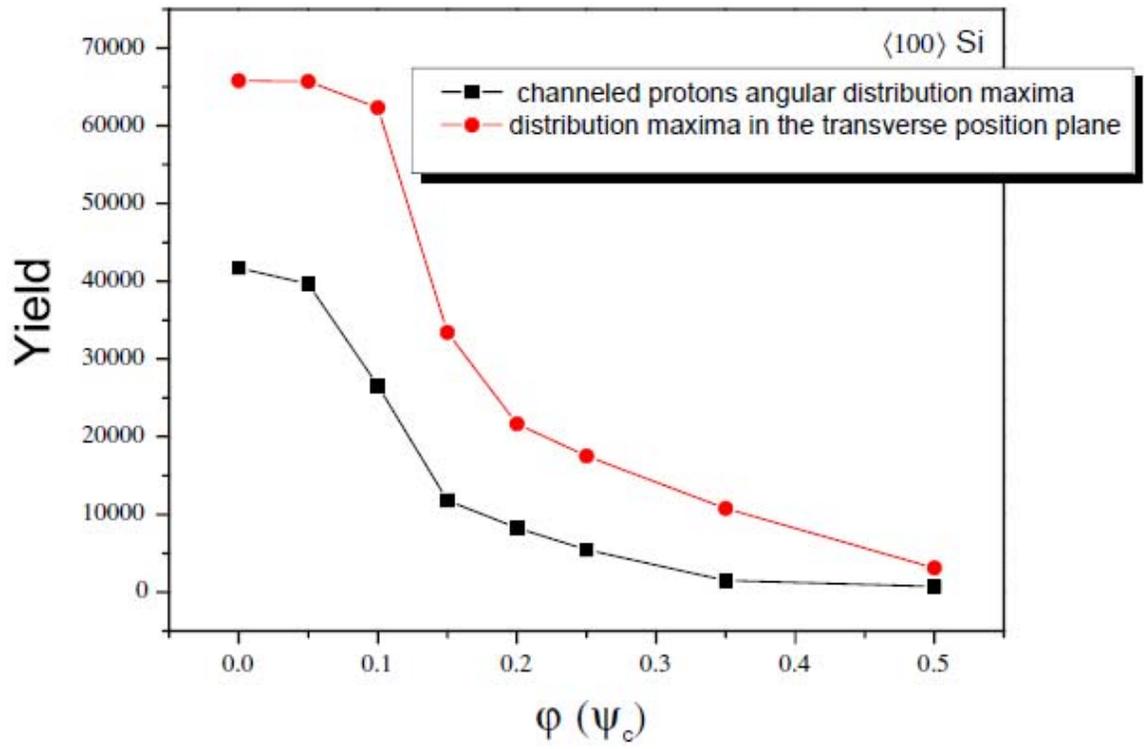

**Figure 4**

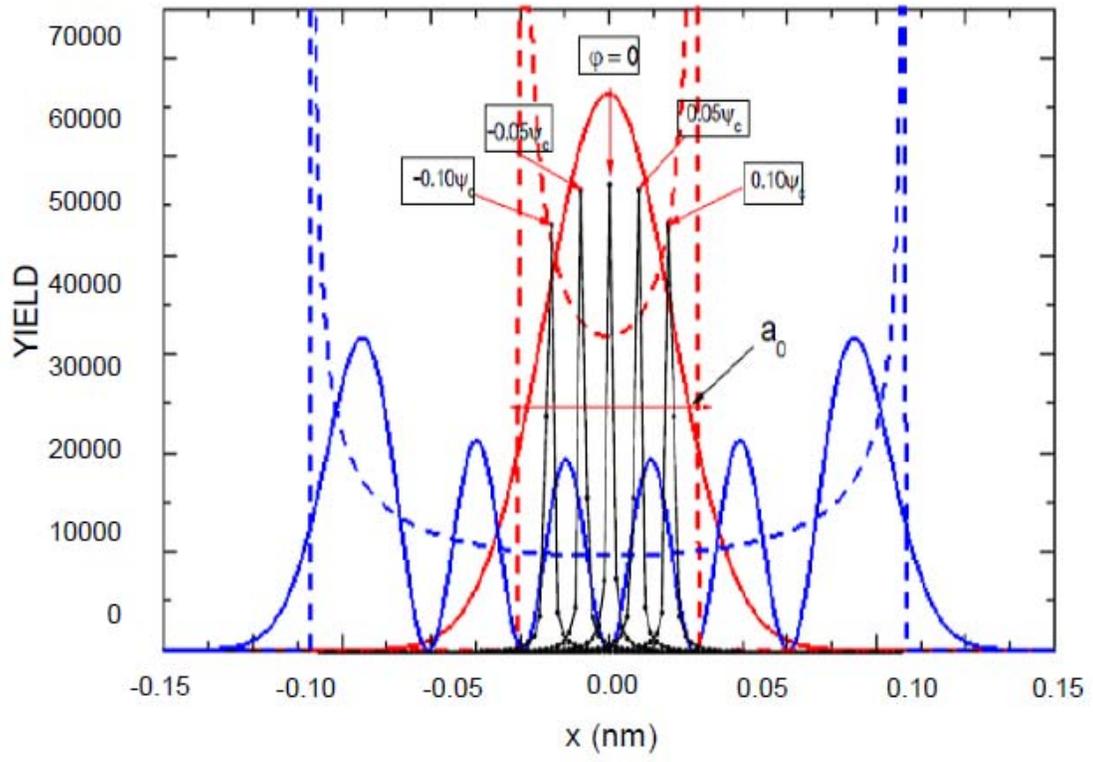

**Figure 5**

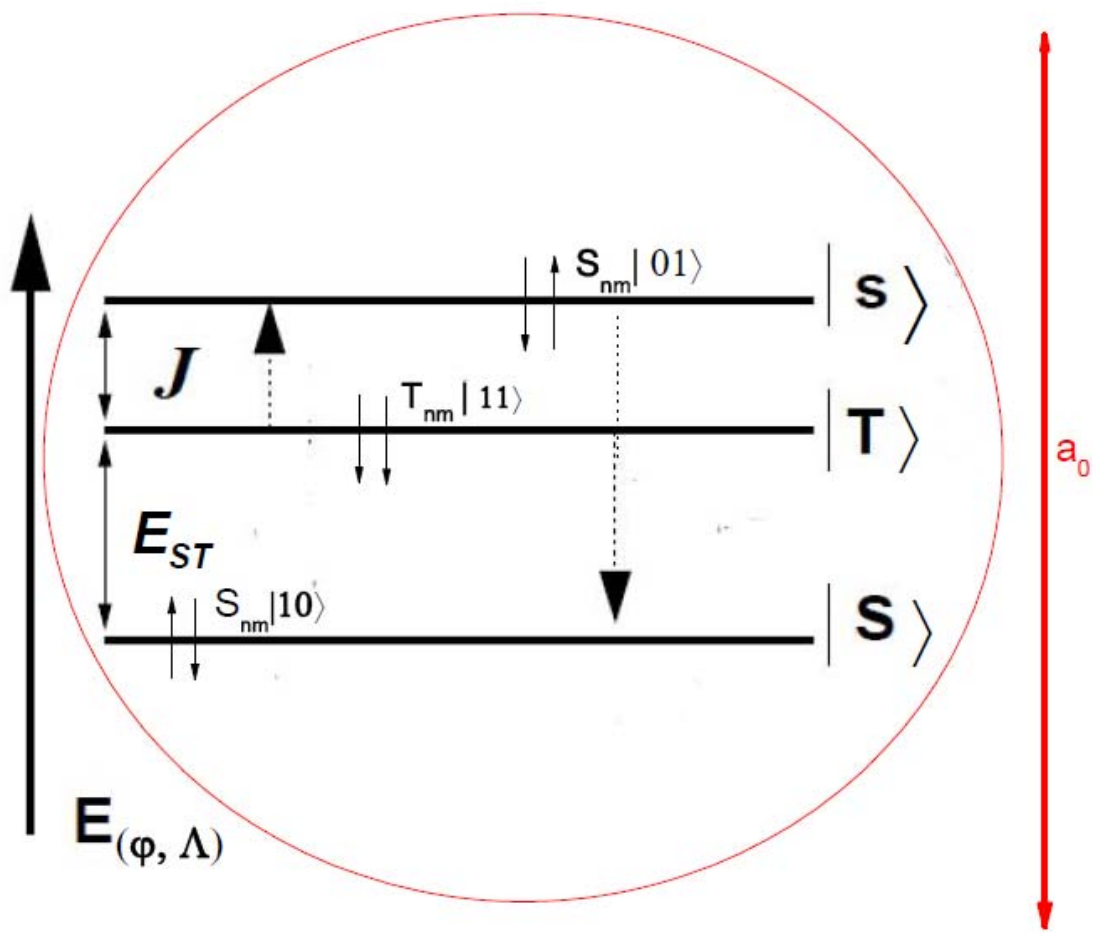

**Figure 6**

a)

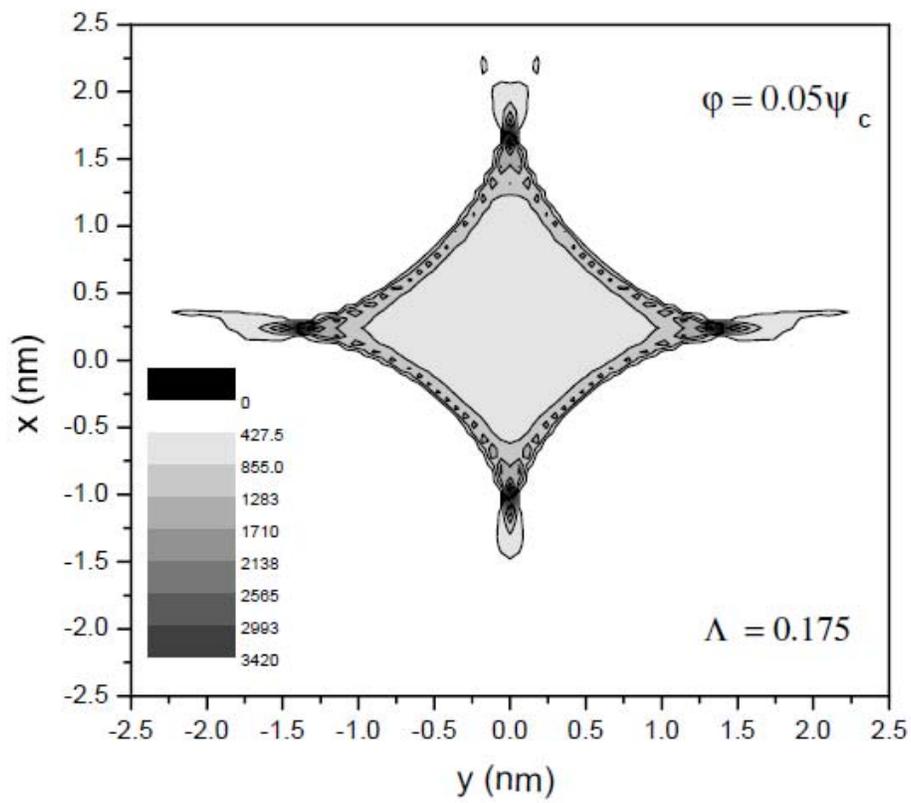

b)

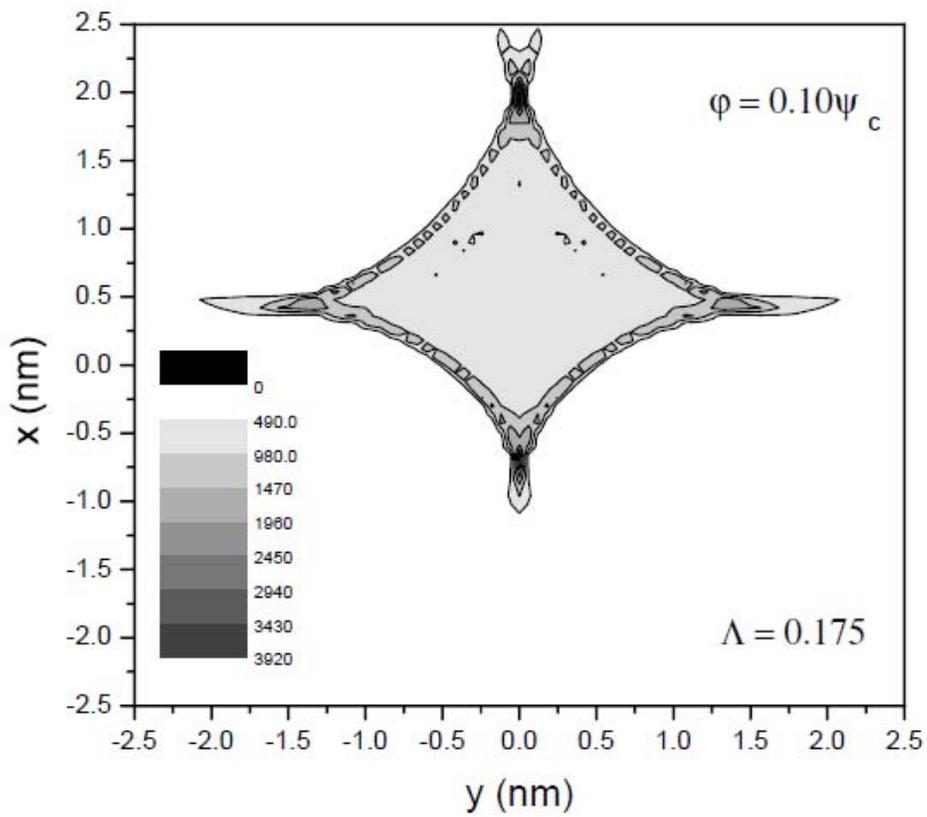

**c)**

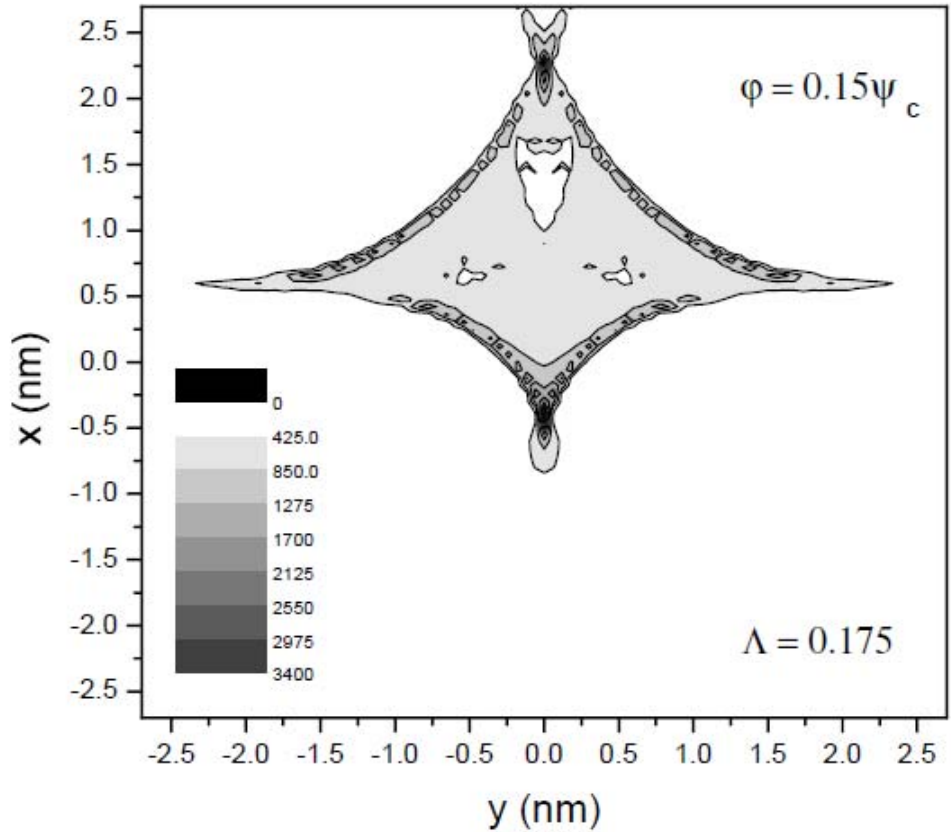